\documentclass[twocolumn,aps,floats,floatfix,groupedaddress]{revtex4-1}
\usepackage{graphicx}
\usepackage{dcolumn}
\usepackage{amsmath}
\usepackage{amsfonts}
\usepackage{amssymb}
\usepackage{epsfig,float,afterpage,wrapfig,psfrag}
\usepackage{natbib}

\newcommand{\beq}{\begin{equation}}
\newcommand{\eeq}{\end{equation}}
\newcommand{\bes}{\begin{subequations}}
\newcommand{\ees}{\end{subequations}}
\newcommand{\bea}{\begin{eqnarray}}
\newcommand{\eea}{\end{eqnarray}}
\newcommand{\ba}{\begin{array}}
\newcommand{\ea}{\end{array}}
\newcommand{\beqn}{\begin{eqnarray*}}
\newcommand{\eeqn}{\end{eqnarray*}}

\newcommand{\f}[2]{\frac{#1}{#2}}

\newcommand{\la}{\langle}
\newcommand{\dg}{\dagger}
\newcommand{\ra}{\rangle}

\def\nn{\nonumber}

\begin{document}

\title{Cascaded two-photon nonlinearity in a one-dimensional waveguide with multiple two-level emitters}

\author{Dibyendu Roy} 
\affiliation{Theoretical Division and Center for Nonlinear Studies, Los Alamos National Laboratory, Los Alamos, New Mexico 87545, USA} 
\begin{abstract}
We propose and theoretically investigate a model to realize cascaded optical nonlinearity with few atoms and photons in one-dimension (1D). The optical nonlinearity in our system is mediated by resonant interactions of photons with two-level emitters, such as atoms or quantum dots in a 1D photonic waveguide. Multi-photon transmission in the waveguide is nonreciprocal when the emitters have different transition energies. Our theory provides a clear physical understanding of the origin of nonreciprocity in the presence of cascaded nonlinearity. We show how various two-photon nonlinear effects including spatial attraction and repulsion between photons, background fluorescence can be tuned by changing the number of emitters and the coupling between emitters (controlled by the separation).
\end{abstract}

\vspace{0.0cm}
\maketitle
 A large range of interesting optical phenomena including all-optical switch \cite{Ironside93, Bosshard95, Baek96}, rectification \cite{Scalora94, Gallo01, Mingaleev02}, squeezing \cite{Kasai97} and bistability \cite{White96}  has been demonstrated employing cascaded optical nonlinearity in macroscopic and mesoscopic systems. These systems are often studied theoretically using classical wave-mechanics \cite{Schiek93, Khurgin97}. On the other side, there has been a fantastic progress in realizing a strong coherent photon-photon nonlinearity at the level of few atoms and photons in various quantum optics set-ups. The efficiency of a single or few atoms to induce strong interactions between propagating photons is indispensable to realize various logic gates for quantum information processing, quantum computation and alternative technologies based on switching and amplification functionalities. One interesting recent proposal to achieve strong coherent photon-photon interactions is by confining photons in reduced dimensions such as, in a one-dimensional (1D) optical waveguide, and coupling these photons with individual emitters in the waveguide \cite{Shen07, Shen07A, Chang07, Zumofen08, Hwang09, Shi09, Roy10, Roy11a, Zheng13, Tsoi08}. Tight confinement of light fields in the waveguide directs majority of the spontaneously emitted light from the emitter into the guided modes, while local interactions at the emitter induce strong photon-photon correlations by preventing multiple occupancy of photons at the emitter. Various nanoscale systems, such as photonic crystal waveguides \cite{Faraon07}, surface plasmon modes of metallic nanowires \cite{Akimov07}, microwave transmission lines \cite{Wallraff04}, optical nanofibers \cite{Dayan08}, semiconductor or diamond nanowires \cite{Claudon10} would act as a 1D continuum for photons. Different two- or multi-level atoms \cite{Dayan08}, molecules \cite{Hwang09}, quantum dots \cite{Akimov07, Claudon10}, superconducting qubits \cite{Wallraff04}, nitrogen-vacancy centres in diamond are used as an emitter to couple with the 1D continuum of photons. 

\begin{figure}[htb]
\includegraphics[width=8.5cm]{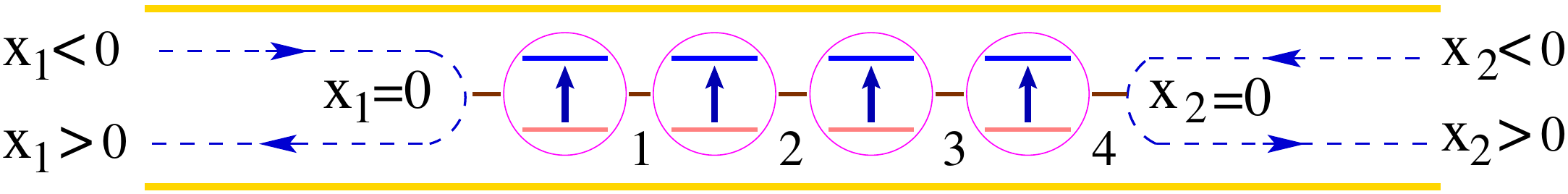}
\caption{A schematic of a chain of two-level emitters embedded in a one-dimensional waveguide, and the propagating free photon modes at the left and the right side of the emitters.}
\label{pl1}
\end{figure}

 Here we propose a microscopic quantum mechanical model to realize cascaded optical nonlinearity with few atoms and photons in 1D. The optical nonlinearity in the proposed model is mediated by resonant interactions of photons with two-level emitters (2LEs) in a 1D photonic waveguide. Multi-photon transmission in the waveguide is nonreciprocal when the emitters have different transition energies. Our theory based on the Bethe-ansatz \cite{Shen07, Shen07A, Roy10, Roy11a} provides a clear physical understanding of the origin of nonreciprocity in the presence of cascaded nonlinearity.  We consider  a chain of 2LEs coupled to propagating free photons in a 1D optical waveguide (see Fig.\ref{pl1}). We assume a small separation between emitters for simplicity, thus we can take an instaneous interaction between emitters. It is so called Markovian limit, where the causal propagation time of photons between two emitters has been neglected \cite{Ficek90, Ficek02}. Two photon correlations in a 1D waveguide with two emitters at arbitrary distance separation has been studied in Ref.\cite{Zheng13} which is based on our previous studies \cite{Dhar08, Roy11b} using the Lippmann-Schwinger scattering theory. Our method in those earlier papers can be further extended for multiple emitters. Here we generalize the recently developed Bethe-ansatz approach \cite{Shen07, Shen07A, Roy10, Roy11a} including open boundary conditions to a chain of multiple emitters coupled to photons. We are able to derive the exact single and two-photon scattering states, and the corresponding photon transmissions in the full system. The Hamiltonian $\mathcal{H}=\mathcal{H}_0+\mathcal{H}_S+\mathcal{H}_C$ in real space within the rotating-wave approximation for the emitter-photon coupling is given by, 
\bea
\mathcal{H}_0&=&-i\int dx ~[v^1_g\tilde{a}^{\dagger}_{1}(x)\partial_x\tilde{a}_{1}(x)+v^2_g\tilde{a}^{\dagger}_{2}(x)\partial_x\tilde{a}_{2}(x)], \nn \\
\mathcal{H}_S&=&\sum_{j=1}^N(\Omega_j-i\f{\gamma_j}{2})|2\ra_{jj}\la2|+\sum_{j=1}^{N-1}J(\sigma_{j,-}\sigma_{j+1,+}+H.c.),\nn\\
\mathcal{H}_C&=&(\tilde{V}_1\sigma_{1,+}\tilde{a}_1(0)+\tilde{V}_2\sigma_{N,+}\tilde{a}_2(0)+H.c.),
\label{Ham0}
\eea
where the first term $\mathcal{H}_0$ represents the propagating photon modes with group velocity $v^j_g$ at the left $(j=1)$ and the right $(j=2)$ side of the chain; the second term $\mathcal{H}_S$ denotes the chain of emitters; and the last term $\mathcal{H}_C$ describes an interaction between the emitters and the free photons. The operators $\tilde{a}^{\dagger}_{1}(x)$ and $\tilde{a}^{\dagger}_{2}(x)$ create a photon at the left and the right of the chain respectively. We set ground state energy of the emitters zero, and $\Omega_j$ is the excited state energy of the $j$th emitter. The exchange coupling between the emitters due to photon mediated instaneous interactions is $J$. Here, $\tilde{V}_{1}~(\tilde{V}_{2})$ is the left (right) emitter-left (right) photons  coupling constant. $\sigma_{j,-}=c_{jg}^{\dagger}c_{je}~(\sigma_{j,+}=c_{je}^{\dagger}c_{jg})$ is a lowering (raising) ladder operator of the $j$th emitter where $c_{jg}^{\dagger} (c_{je}^{\dagger})$ is a creation operator of the ground state $|1\ra_j$ (excited state $|2\ra_j$) of the $j$th emitter. There would be a loss of spontaneously emitted photons to the non-guided modes. It is usually incorporated by including an imaginary term -$i\gamma_j/2$ in the energy of the excited emitter states within the quantum jump picture. In our study the 1D features of scattering come from the interference of the spontaneously emitted photons in guided modes with the incident photons in the guided modes. A finite value of $\gamma_j (\sim \Gamma)$ reduces the 1D scattering features by reducing the spontaneously emitted photons in the guided modes. However recent studies \cite{MangaRao07, Lund08} have shown a significant control over the loss of spontaneously emitted photons to the non-guided modes. Therefore, we set $\gamma_j=0$ in all the plots. Next we scale the free photon operators to absorb the group velocity, and redefine $\sqrt{v^j_g}\tilde{a}_{j}(x)\equiv a_j(x)$, $\tilde{V}_j/\sqrt{v^j_g} \equiv V_j$. Therefore we rewrite $\mathcal{H}_0=-i\int dx \sum_{j=1,2}a^{\dagger}_{j}(x)\partial_xa_{j}(x)$ and $\mathcal{H}_C=V_1\sigma_{1,+}a_1(0)+V_2\sigma_{N,+}a_2(0)+H.c.$. Hereafter we always consider the transformed Hamiltonian. The coupling of the emitters to the waveguide fields $\Gamma_{1,2}=V_{1,2}^2$ are related to the spontaneous decay rate of the emitters by $1/\tau_{1,2}=\Gamma_{1,2}$. We describe the results of a minimal model of the chain, namely a chain of two emitters, $N=2$ in the main text, and include the results for a chain of three emitters, $N=3$ in the Appendix (see Supplementary Information of the published version, D. Roy, Sci. Rep. {\bf 3}, 2337 (2013)). It is possible to extend our generalized approach to calculate the scattering state and the transmission for three or more photons and for arbitrary $N$. A chain of coupled $N$ 2LEs can be mapped to a single emitter with $2^N$ energy levels. However, the present method works better in the representation of coupled $N$ 2LEs as it is easier to write down an ansatz for the full scattering state with the 2LEs being either in the ground state or in the excited state.

{\large {\bf Results}}\\
{\bf Single-photon dynamics:}  We take an incident photon with wavevector $k$ (and energy $E_k=k$) injected from the left of the emitters. In our approach the wavefunction at $x<0$ describes the full system before scattering of photons from the emitters, and the wavefunction at $x>0$ characterizes the system after scattering.  The single-photon incident state is $|\psi^1_{\rm in}\ra$, and the outgoing wavefunction is $|\psi^1_{\rm out}\ra$. $|\psi^1_{\rm in}\ra=\f{1}{\sqrt{2\pi}}\int dx~ e^{ikx}a^{\dagger}_{1}(x)|0,1,1\ra$ and $|\psi^1_{\rm out}\ra=\f{1}{\sqrt{2\pi}}\int dx \big[(\phi^1_k(x)a^{\dagger}_{1}(x)+ \phi^2_k(x)a^{\dagger}_{2}(x))|0,1,1\ra+\delta(x)(e^1_k|0,2,1\ra+e^2_k|0,1,2\ra)\big]$, where $|n,l_1,m_2\ra$ denotes the state of the full system with $n$ number of photons in the waveguide, the left emitter in $l_1$ state and the right emitter in $m_2$ state. The amplitude of the $j$th emitter in the excited state is $e^j_k$. We find different amplitudes in $|\psi^1_{\rm out}\ra$ by solving a set of linear differential equations obtained from the stationary single-photon Schr{\"o}dinger equation, $\mathcal{H}|\psi^1_{\rm out}\ra=E_k|\psi^1_{\rm out}\ra$ (see Supplementary Information of the published version, D. Roy, Sci. Rep. {\bf 3}, 2337 (2013)). $e^2_k=Je^1_k/(E_k-\Omega_2+i\Gamma_2/2),~ e^1_k=V_1/(\chi+i\Gamma_1/2),~\phi^1_k(x)=e^{ikx}~\theta(-x)+r^1_ke^{ikx}~\theta(x),~\phi^2_k(x)=t^1_ke^{ikx}\theta(x)$,
\begin{figure}
\includegraphics[width=8.5cm]{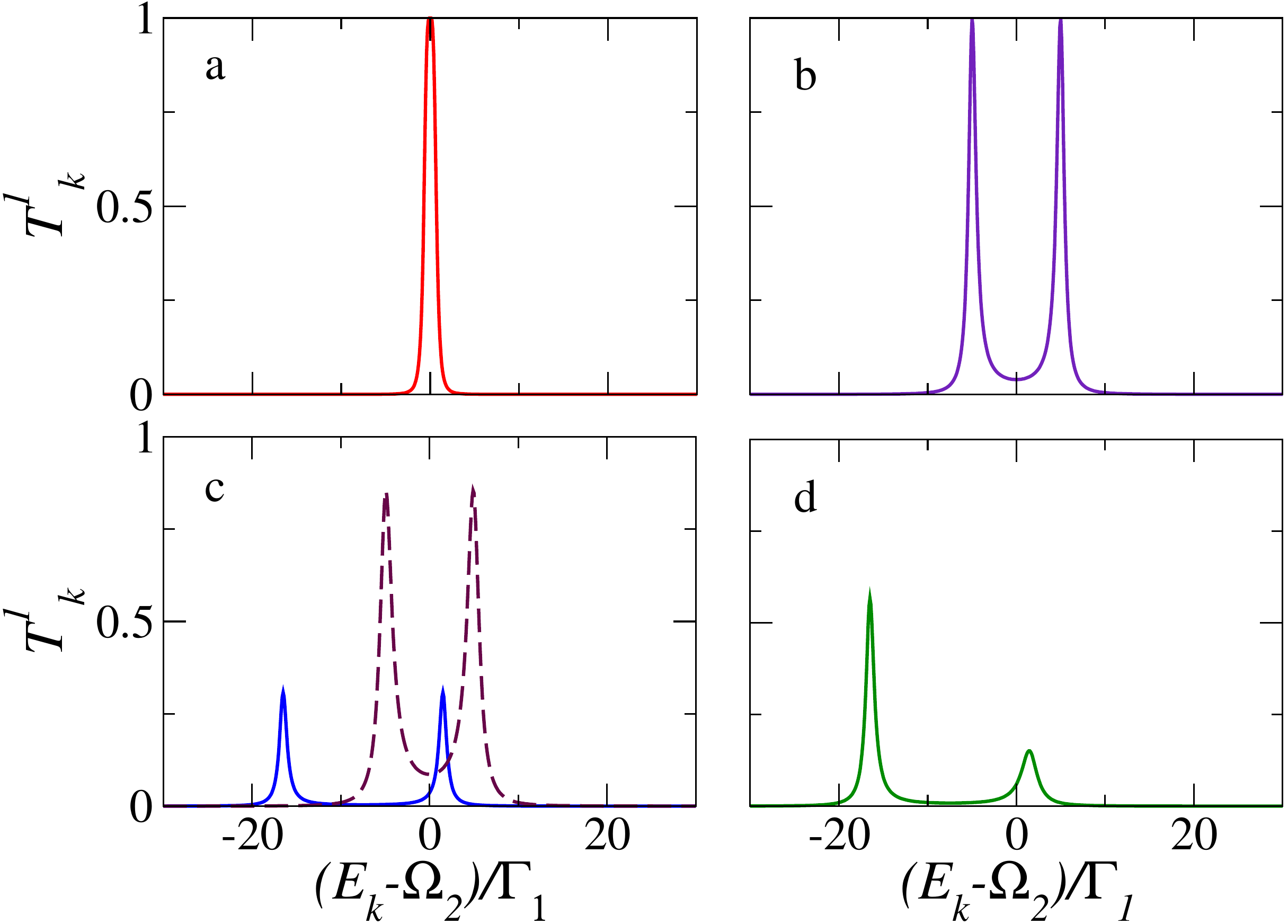}
\caption{Single photon transmission $T^1_k$ through two emitters versus scaled energy $(E_k-\Omega_2)/\Gamma_1$ of an incident photon. The parameters are (a) $\Omega_1=\Omega_2$, $\Gamma_2=\Gamma_1$, $J=\Gamma_1/2$, (b) $\Omega_1=\Omega_2$, $\Gamma_2=\Gamma_1$, $J=5\Gamma_1$, (c) $\Omega_1=\Omega_2-15\Gamma_1$, $\Gamma_2=\Gamma_1$, $J=5\Gamma_1$ (blue full curve), $\Omega_1=\Omega_2$, $\Gamma_2=2.25\Gamma_1$, $J=5\Gamma_1$ (maroon dash curve) and (d) $\Omega_1=\Omega_2-15\Gamma_1$, $\Gamma_2=2.25\Gamma_1$, $J=5\Gamma_1$.}
\label{pl2}
\end{figure}
where the single-photon transmission and reflection amplitudes are respectively $t^1_k=\f{-iV_1V_2J}{(\chi+i\Gamma_1/2)(E_k-\Omega_2+i\Gamma_2/2)}$ and $r^1_k=\f{\chi-i\Gamma_1/2}{\chi+i\Gamma_1/2}$, and $\chi=E_k-\Omega_1-J^2/(E_k-\Omega_2+i\Gamma_2/2)$, $\Gamma_i=V^2_i$. For $\Gamma_1=\Gamma_2=\Gamma$, and two identical emitters $\Omega_1=\Omega_2=\Omega$, the transmission coefficient $T^1_k=|t^1_k|^2$ becomes one at $E_k=\Omega$ and $J=\Gamma/2$ (see Fig.\ref{pl2}(a)), and the corresponding reflection coefficient is zero. We call it a single peak resonance (SPR). We plot $T^1_k$ through two emitters in Fig.\ref{pl2} for different parameter sets. The transmission curve always has two peaks except at the SPR. The single-photon transmission curve becomes asymmetric in shape only when both $\Omega_1\ne \Omega_2$ and $\Gamma_1\ne \Gamma_2$. When the coupling $J$ between two emitters is relatively weak, i.e., $J\le \Gamma_1,\Gamma_2$,  the resonance peaks appear near the transition energy $\Omega_j$ of the emitters. However for a stronger coupling between emitters ($J>\Gamma_1,\Gamma_2$), the resonant peaks appear at modified energies which can be calculated by diagonalizing the isolated chain of emitters. The transmission of a single photon in our system can be detected by analyzing the temporal correlations of photons at the exit of the waveguide using single photon detectors for optical frequencies and linear detectors for microwave frequencies. 
\begin{figure*}[htb]
\includegraphics[width=16.5cm]{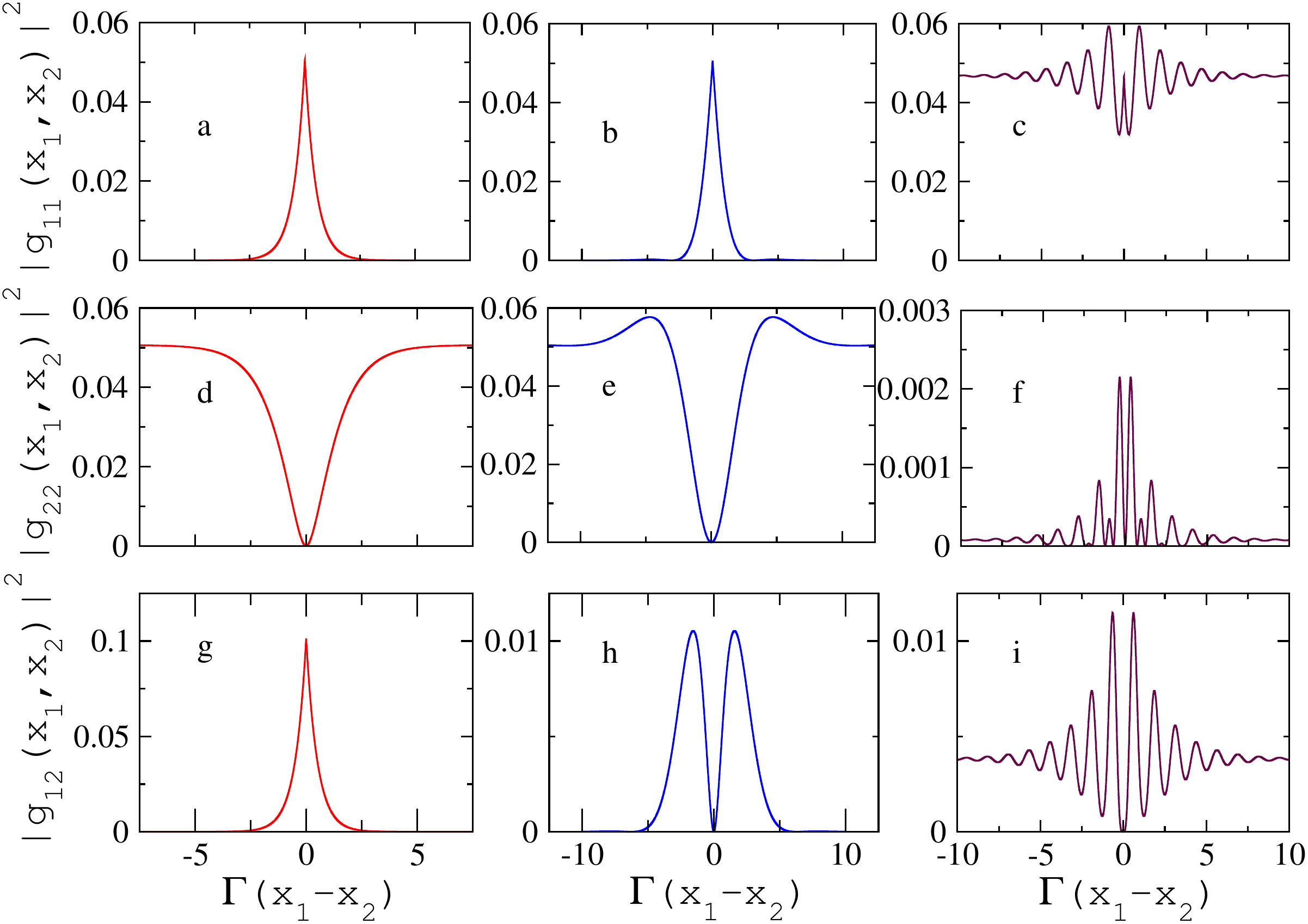}
\caption{Two-photon correlations $|g_{11}(x_1,x_2)|^2,~|g_{22}(x_1,x_2)|^2,~|g_{12}(x_1,x_2)|^2$ for one emitter (first column) and two identical emitters (middle and last columns). Here, $\Omega_1=\Omega_2=\Omega$ and $\Gamma_1=\Gamma_2=\Gamma$ in all the plots for two emitters. The parameters are, $E_{k_1}=E_{k_2}=\Omega$. The coupling $J=\Gamma/2$ for the middle column and $J=5\Gamma$ for the last column.}
\label{2PhCor}
\end{figure*}

{\bf Two-photon dynamics:}  Two-photon scattering states for a single emitter coupled to photons are evaluated by introducing an even-odd transformation of the incident photons \cite{Shen07, Roy10, Roy11a}. This transformation simplifies the calculation as photons in the even mode is only coupled to the emitter. Similar transformation of the free photon modes is not useful for this problem as two different emitters are coupled to different photon modes. Instead we here derive the two-photon scattering states using the original photon modes. This procedure also helps us later to derive scattering states for a chain of multiple emitters. The two-photon incoming state $|\psi^2_{\rm in}\ra$ for two injected photons from the left is given by, $|\psi^2_{\rm in}\ra=\int dx_1 dx_2 \phi_{\bf k}(x_1,x_2)\f{1}{\sqrt{2}}a^{\dg}_1(x_1)a^{\dg}_1(x_2)|0,1,1\ra$ where $\phi_{\bf k}(x_1,x_2)=(e^{ik_1x_1+ik_2x_2}+e^{ik_1x_2+ik_2x_1})/2\pi \sqrt{2}$ with the incident wave vector ${\bf k}=(k_1,k_2)$. The total energy of two incident photons $E_{\bf k}=k_1+k_2$. We write a general two-photon scattering state $|\psi^2_{\rm out}\ra$ using the operators of free photon modes and emitters.
\bea
&&|\psi^2_{\rm out}\ra=\int dx_1dx_2\Big[\big\{g_{11}(x_1,x_2)\f{1}{\sqrt{2}}a^{\dg}_1(x_1)a^{\dg}_1(x_2)\nn\\&&+e^1_1(x_1)\delta(x_2)a^{\dg}_1(x_1)\sigma_{1+}+e^2_1(x_1)\delta(x_2)a^{\dg}_1(x_1)\sigma_{2+}\nn \\&&+e_{12}\delta(x_1)\delta(x_2)\sigma_{1+}\sigma_{2+}\big\}+\big\{g_{12}(x_1;x_2)a^{\dg}_1(x_1)a^{\dg}_2(x_2)\nn\\&&+e^1_2(x_2)\delta(x_1)a^{\dg}_2(x_2)\sigma_{1+}+e^2_2(x_2)\delta(x_1)a^{\dg}_2(x_2)\sigma_{2+}\big\}\nn \\ &&+g_{22}(x_1,x_2)\f{1}{\sqrt{2}}a^{\dg}_2(x_1)a^{\dg}_2(x_2)\Big] |0,1,1\ra,
\label{2LEwavefn}
\eea 
where $g_{11}(x_1,x_2) \equiv g_{11}(x_2,x_1)$ and $g_{22}(x_1,x_2) \equiv g_{22}(x_2,x_1)$ for the Bose statistics of photons.  The amplitudes $g_{11}(x_1,x_2),~ g_{22}(x_1,x_2)$, and $g_{12}(x_1;x_2)$ denote outgoing two-photon wavefunctions, in which either both photons are reflected or transmitted, or one photon is transmitted while the other is reflected. 
Here $e^1_1(x),e^2_1(x)~(e^1_2(x),e^2_2(x))$ are the amplitudes of one photon in the left (right) side of the chain while respectively the left or the right emitter in the excited state. The amplitude of both the emitters in the excited state is $e_{12}$.  We evaluate various amplitudes in Eq.(\ref{2LEwavefn}) using the two-photon Schr{\"o}dinger equation, $\mathcal{H}|\psi^2_{\rm out}\ra=E_{\bf k}|\psi^2_{\rm out}\ra$ (see Supplementary Information of the published version, D. Roy, Sci. Rep. {\bf 3}, 2337 (2013)). 

The solutions of two-photon wave-functions $g_{11}(x_1,x_2),~ g_{22}(x_1,x_2)$ and $g_{12}(x_1;x_2)$ contain an inelastic contribution from a {\it two-photon bound state} which arises due to exchange of energy and momentum between two scattered photons. We call them by {\it bound states} as these terms fall rapidly with increasing separation, $|x_1-x_2|\equiv |x|$ between two photons as shown in Eqs.\ref{bound}.  
These {\it bound states} are the origin of background fluorescence which can be conceived as an inelastic scattering of one photon from a composite transient object formed by the emitter absorbing  the other photon. We derive form of the two-photon wavefunctions at the SPR of the identical emitters. The wavefunctions at this special point behave as 
\bea
g_{11}(x_1,x_2)&=&-\f{1}{\sqrt{2}\pi}e^{iE_{\bf k}x_c}e^{-\Gamma|x|/2}\cos(\Gamma|x|/2),\nn\\
g_{22}(x_1,x_2)&=&- \f{1}{\sqrt{2}\pi}e^{iE_{\bf k}x_c}(1-e^{-\Gamma|x|/2}\cos(\Gamma|x|/2)),\nn \\
g_{12}(x_1,x_2)&=&\f{i}{\pi}e^{iE_{\bf k}x_c} e^{-\Gamma|x|/2} \sin (\Gamma x/2),\label{bound}
\eea
with $x_c=(x_1+x_2)/2$. These functions for a direct coupled 2LE-photons system in a 1D waveguide at the single-photon resonance are given by $\bar{g}_{11}(x_1,x_2)=-e^{iE_{\bf k}x_c}e^{-\Gamma|x|/2}/(\sqrt{2}\pi),~\bar{g}_{22}(x_1,x_2)=-e^{iE_{\bf k}x_c}(1-e^{-\Gamma|x|/2})/(\sqrt{2}\pi),~\bar{g}_{12}(x_1;x_2)=-e^{iE_{\bf k}x_c}e^{-\Gamma|x|/2}/\pi$. Thus, there is an extra sinusoidal oscillation for the two emitters compared to the single emitter. These are shown in the first (one emitter) and the second column (two emitters) of Fig.\ref{2PhCor}. The sinusoidal oscillation at the SPR creates a drastic change in the form of $g_{12}(x_1,x_2)$ for the two emitters from that of a single. The {\it anti-bunching} of two transmitted photons in the direct coupled emitters model occurs because two photons can not be emitted simultaneously by the right (second) emitter, and the transmitted photons in the right side of the waveguide are solely due to emission from the right emitter. The {\it anti-bunching} is considered as spatial repulsion between photons while the {\it bunching}, when two photons are prone to come simultaneously is known as spatial attraction between photons. The {\it bunching} of two reflected photons is shown in Fig.\ref{2PhCor}(a),(b). The sinusoidal oscillation in $g_{11}(x_1,x_2)$ for the two emitters is disguised by fast exponential decay in Fig.\ref{2PhCor}(b). In the  last column of Fig.\ref{2PhCor}, we show correlations of two scattered photons for the two emitters by increasing the coupling $J$ between the emitters. Multiple scattering of photons between the emitters creates many oscillations in the two-photon wavefunctions away from the SPR. The number of oscillations in the two-photon correlations increases with a stronger coupling $J$ between emitters. Specially the amplitude of two transmitted photons reduces by one order of magnitude because the single photon transmission is much reduced for $E_{k_1}=E_{k_2}=\Omega$ at $J=5\Gamma$. A possible experimental set-up to measure the two-photon correlations of a single emitter coupled to photon modes in a waveguide has been proposed in Ref.\cite{Shen07A}  by placing a beam splitter at the ends of the 1D waveguide with a single-photon counter on each output arm of the
beam splitter. This set-up can be used to measure the two-photon correlations for multiple emitters.

\begin{figure}[htb]
\includegraphics[width=8.75cm]{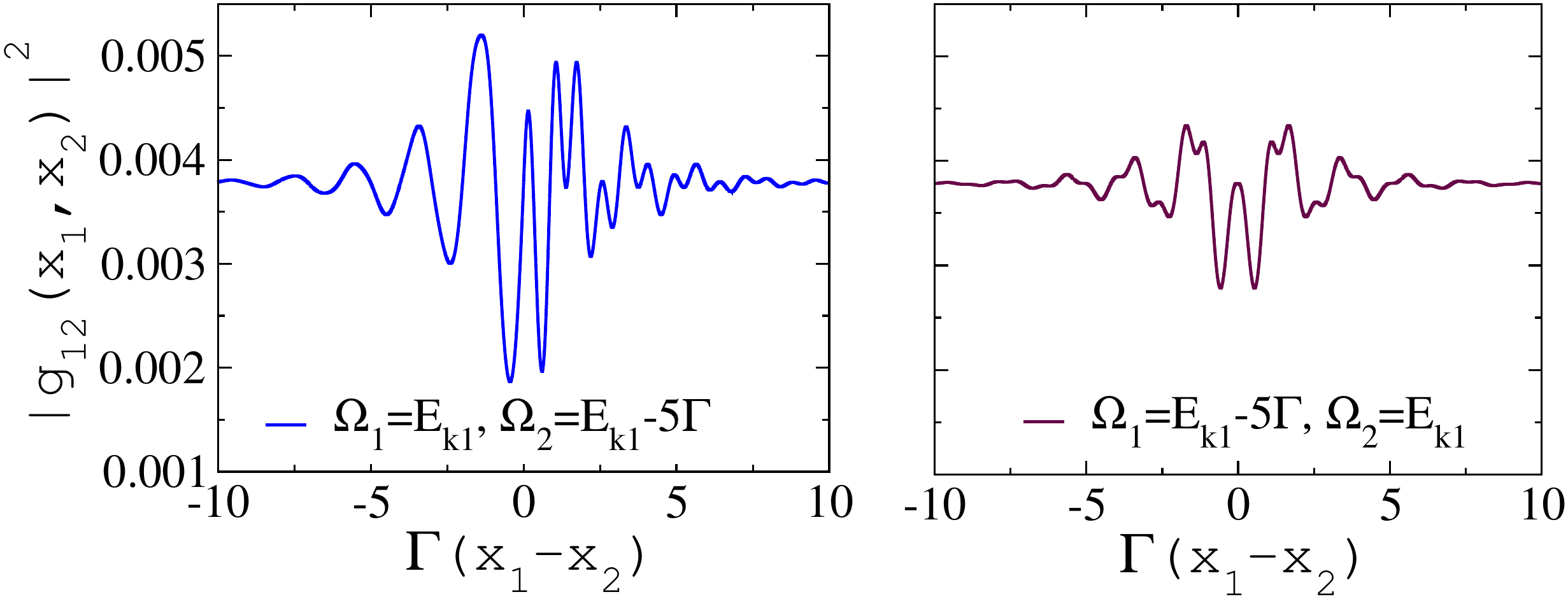}
\caption{The figure illustrates non-reciprocity in two different emitters. Correlation $|g_{12}(x_1,x_2)|^2$ of one reflected and one transmitted photons in two different emitters for incident photons from the left at $E_{k_1}=E_{k_2}$ and $\Gamma_1=\Gamma_2=\Gamma,~J=5\Gamma$.}
\label{2PhCorA}
\end{figure}

 {\bf Optical nonreciprocity:} We define a photon current operator, $\hat{I}=i[\mathcal{H}, N_1-N_2]/2$ where $N_1~(N_2)$ is an operator of total number of photons in the left (right) side of the emitter chain. Therefore, $\hat{I}=i(V_1a^{\dg}_1(0)\sigma_{1,-}-V_2a^{\dg}_2(0)\sigma_{2,-}-H.c.)/2$. We take expectation of $\hat{I}$ in $|\psi^1_{\rm out}\ra$ and $|\psi^2_{\rm out}\ra$ to derive the single and two-photon current. The single-photon current for a photon coming from the left ($j=1$) or right ($j=2$) is given by $I(j;k)=\la \psi^1_{\rm out}|I|\psi^1_{\rm out}\ra=|t^1_k|^2/(2\pi)$. We find, $I(1;k)=I(2;k)$ irrespective of the values of couplings and transition energies, including $V_1\ne V_2$ and $\Omega_1\ne \Omega_2$. Thus the single photon transmission is always reciprocal across the two emitters. The two-photon current $I(1,k_1,k_2)$ for two incident photons from the left has two parts, one $I_0(k_1,k_2)$ is a contribution of two noninteracting photons, and the other $\delta I(1,k_1,k_2)$ is a change in the two-photon current due to photon-photon interactions. 
\bea
&&I(1,k_1,k_2)=I_0(k_1,k_2)+\delta I(1,k_1,k_2),~~{\rm where}\nn\\
&&I_0(k_1,k_2)=\f{\mathcal{L}}{4\pi^2}(|t^1_{k_1}|^2+ |t^1_{k_2}|^2 + 2 |t^1_{k_1}|^2\delta_{k_1,k_2}).
\eea
Here $\mathcal{L}$ denotes the length of the 1D waveguide. Again, $I_0(k_1,k_2)$ is the same for incident photons from the left or the right. The magnitude of interaction induced current change $\delta I$ is different for incident photons from the left or the right side of the emitters when either $V_1\ne V_2$ or $\Omega_1 \ne \Omega_2$ or $V_1\ne V_2$ and $\Omega_1 \ne \Omega_2$. Therefore, two-photon transmission across the emitters is nonreciprocal whenever parity (mirror symmetry) of the system is broken. To understand the physical origin of such interesting nonreciprocity, we check nature of the two-photon wavefunctions by interchanging $\Omega_1,~\Omega_2$ or $V_1,~V_2$. Because, if the magnitude of $I(1,k_1,k_2)$ changes by an interchange of $\Omega_1,~\Omega_2$ or $V_1,~V_2$, that is surely equivalent to asymmetric two-photon current in this system. For two incident photons from the left, the form of $g_{11}(x_1,x_2)$ and $g_{12}(x_1,x_2)$ is transformed by an interchange of $\Omega_1,~\Omega_2$ (check Fig.\ref{2PhCorA}) or $V_1,~V_2$ while $g_{22}(x_1,x_2)$ remains the same. Contribution in the two-photon current comes both from $g_{12}(x_1,x_2)$ and $g_{22}(x_1,x_2)$. Thus, a difference in $g_{12}(x_1,x_2)$ by an interchange of the transition energies or the couplings is the reason for nonreciprocity in our model. Next we ask why there is a change in the one reflected and one transmitted wavefunction but not in the both transmitted wavefunction when we interchange $\Omega_1,~\Omega_2$ or $V_1,~V_2$. The interaction of an emitter with multiple photons creates an effective nonlinear interaction between photons. Two transmitted photons see the full part of the nonlinear interaction created by the two different emitters while one reflected-one transmitted photons and two reflected photons see only part of the nonlinear interaction depending on the direction of the incoming photons.  

{\large{\bf Discussion}}

The present microscopic model with just two different 2LEs is the smallest physical system showing cascaded optical nonlinearity. We provide a fully quantum mechanical description to understand the response of individual atoms to an applied weak light field. Recent experiments have demonstrated a large optical phase shift in light scattered by a single isolated atom to validate a microscopic model that underpins the macroscopic phenomenon of the refractive index \cite{Jechow12}. Therefore, our simple model provides a link in the origin of the refractive index from single atoms to bulk nonlinear medium.  There are several different proposals for realizing optical nonreciprocity or diode using various mechanisms, such as, magneto-optic effect, macroscopic and mesoscopic optical nonlinearity \cite{Fan12}, indirect inter-band photonic transitions \cite{Yu09}, opto-acoustic effect \cite{Kang11}. Our proposed optical diode works at low intensity of light in the fully-quantum regime compared to most previous proposals in the classical regime, and may have potential applications to build quantum logic gates  for optical quantum information processing and quantum computation. The advantage of the present set-up of an optical diode compared to the one in Ref.\cite{Roy10} is that we do not need asymmetric emitter-photon couplings to generate nonreciprocal photon transmission. It is experimentally challenging to create such asymmetric emitter-photon coupling with a single emitter. Here we show that the nonreciprocal photon transmission can be generated at few-photon level even for symmetric coupling with two different emitters in a waveguide. The present set-up is much easier to realize in experiments. In Ref.\cite{Yao09} a related system of coupled emitters placed in cavities which are spatially separated with a waveguide has been studied. The amount of nonreciprocity in optical transmission is expected to be higher in the  waveguide-cavity systems than in the waveguide due to a stronger light-matter coupling in the waveguide-cavity systems. However we need further study along this direction.

{\large \bf Acknowledgments} \\
The support of the U.S. Department of Energy through LANL/LDRD Program for this work is gratefully acknowledged.

\end{document}